\pdfoutput=1

\documentclass[aps,prb,twocolumn,amsmath,superscriptaddress]{revtex4}
\usepackage{graphicx}
\usepackage{color}

\begin{document}

\title{Optical control of the spin state of two Mn atoms in a quantum dot}

\author{L. Besombes}
\email{lucien.besombes@grenoble.cnrs.fr}
\affiliation{CEA-CNRS group "Nanophysique et
semiconducteurs", Institut N\'eel, CNRS \& Universit\'e
Joseph Fourier, 25 avenue des Martyrs, 38042 Grenoble,
France}

\author{C.L. Cao}
\affiliation{CEA-CNRS group "Nanophysique et
semiconducteurs", Institut N\'eel, CNRS \& Universit\'e
Joseph Fourier, 25 avenue des Martyrs, 38042 Grenoble,
France}\affiliation{Departamento de F\'{\i}sica Aplicada,
Universidad de Alicante, San Vicente del Raspeig, 03690
Spain}

\author{S. Jamet} \affiliation{CEA-CNRS group
"Nanophysique et semiconducteurs", Institut N\'eel, CNRS \&
Universit\'e Joseph Fourier, 25 avenue des Martyrs, 38042
Grenoble, France}

\author{H. Boukari} \affiliation{CEA-CNRS group
"Nanophysique et semiconducteurs", Institut N\'eel, CNRS \&
Universit\'e Joseph Fourier, 25 avenue des Martyrs, 38042
Grenoble, France}

\author{J. Fern\'andez-Rossier} \affiliation{Departamento
de F\'{\i}sica Aplicada, Universidad de Alicante, San
Vicente del Raspeig, 03690 Spain}
\affiliation{International Iberian Nanotechnology
Laboratory, Av. Mestre Jos\'e Veiga, 4715-330 Braga,
Portugal}

\date{\today}

\begin{abstract}

We report on the optical spectroscopy  of the spin  of two
magnetic atoms (Mn) embedded in an individual quantum dot
interacting with either a single electron, a single exciton
and single trion. As a result of their interaction to a
common entity, the Mn spins become correlated. The dynamics
of this process is probed by time resolved spectroscopy,
that permits to determine the optical orientation time in
the range of a few tens of $ns$. In addition, we show  that
the energy of the collective spin states of the two Mn
atoms can be tuned through the optical Stark effect induced
by a resonant laser field.
\end{abstract}

\maketitle

Single atom  quantum devices\cite{Solotronics}, i.e.,
systems  whose macroscopic properties depend on the quantum
state of a single atom, have been demonstrated in several
systems, like single donor Silicon
transistor\cite{single-atom-transistor2012}, single Mn
doped quantum dot (QD)\cite{Besombes2004}, a single
magnetic atom in a surface\cite{Hirjibehedin_2007} or a
single NV center in nano diamond\cite{NV1}.    The
manipulation of the spin state of a single atom has been
shown both by optical pumping\cite{LeGall2009,Goryca2009}
and in transport experiments\cite{Loth_2010}.  These
systems permit to test matter at a fundamental scale and,
in some instances, already  have practical applications
like room temperature magnetometry with nanometer
resolution\cite{NV2}. A controlled upscale of this primary
units would permit to test new physical phenomena and to
find new applications.  In this regard, the study of chains
of a few magnetic atoms deposited on a metal is already
giving promising results along this line\cite{Loth2012}.
Here we report on the first step in that direction in the
case of Mn atoms in semiconductor QDs.

When Mn atoms are included in a II-VI semiconductor QD
(CdTe in ZnTe)\cite{Wojnar2011}, the spin of the optically
created electron-hole pair (exciton) interacts with the
5{\it d} electrons of the Mn (total spin S=5/2). In the
case of a singly Mn doped QD, this leads to a splitting of
the once simple photoluminescence (PL) spectrum of an
individual QD into six (2S+1) components
\cite{Besombes2004}.

In contrast to the case of magnetic atoms in a metal,
semiconductors afford the unique opportunity of controlling
the exchange interaction between distant magnetic dopants
by varying the carrier density of the host. This has been
shown in quantum wells\cite{Boukari2002} but in quantum
dots the effect is expected to be stronger and to  take
advantage of the discreteness of the charge
addition\cite{Fernandez2004,Qu2005}. Thus,  in a neutral
dot containing 2 Mn atoms, the spins are only coupled via a
very short range supercharge, which would be only relevant
for first or second neighbors \cite{Furdyna1988,Qu2006}.
The injection of a single electron, whose wave function is
spread along the entire dot, couples the 2 Mn and the
electron spin ferromagnetically, resulting in a ground
state with $S=11/2$. This contrasts with the addition of a
single exciton on the neutral dot, for which the Mn spins
also couple ferromagnetically, but the  strong spin-orbit
coupling of the hole breaks spin rotational invariance. The
addition of an exciton on the negatively charged dot, puts
the 2 electrons in a singlet state so that the  2 Mn
interacting with a single hole.

Here we show how we can address and investigate the
magnetic properties of 2 Mn atoms which are coupled to
optically injected carriers. The fine structure of a
confined exciton (X) and negatively charged exciton ($X^-$)
in the exchange field of the 2 Mn atoms are analyzed and
modeled in detail. We show that the 2 Mn spins can be
optically oriented by the injection of spin polarized
carriers in a few $ns$. Finally, we demonstrate that the
energy of the spin states of the 2 Mn atoms can be tuned by
a resonant laser field.

\begin{figure}[hbt]
\includegraphics[width=3.3in]{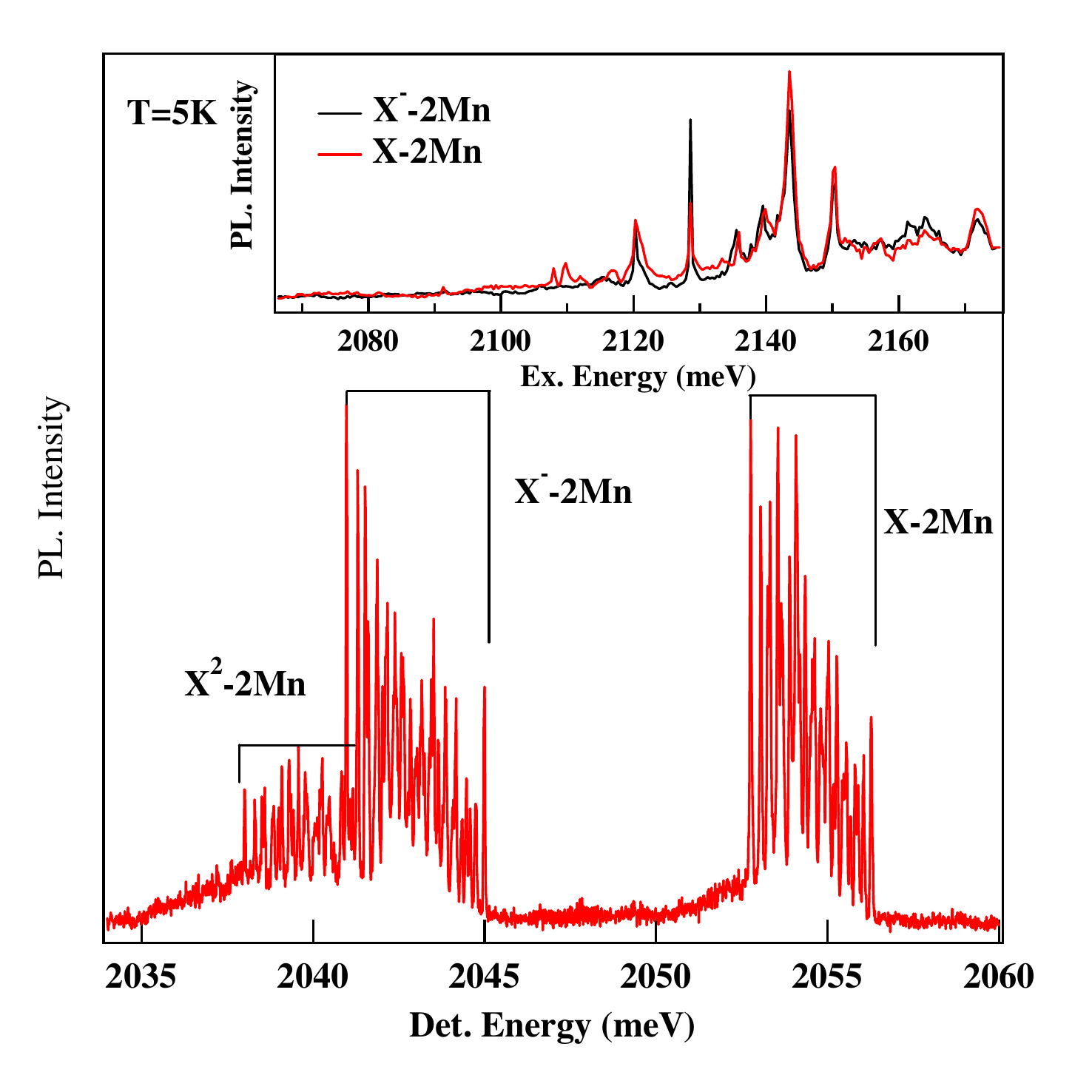}
\caption{PL spectrum showing the coexistence of the neutral
exciton (X-2Mn), the negatively charged exciton
(X$^{-}$-2Mn) and biexciton (X$^{2}$-2Mn) exchanged coupled
to 2 Mn atoms in a CdTe/ZnTe QD. Inset: PL excitation
spectra detected on the low energy line of X-2Mn and
X$^{-}$-2Mn.} \label{fig1}
\end{figure}

\begin{figure}[hbt]
\includegraphics[width=3.3in]{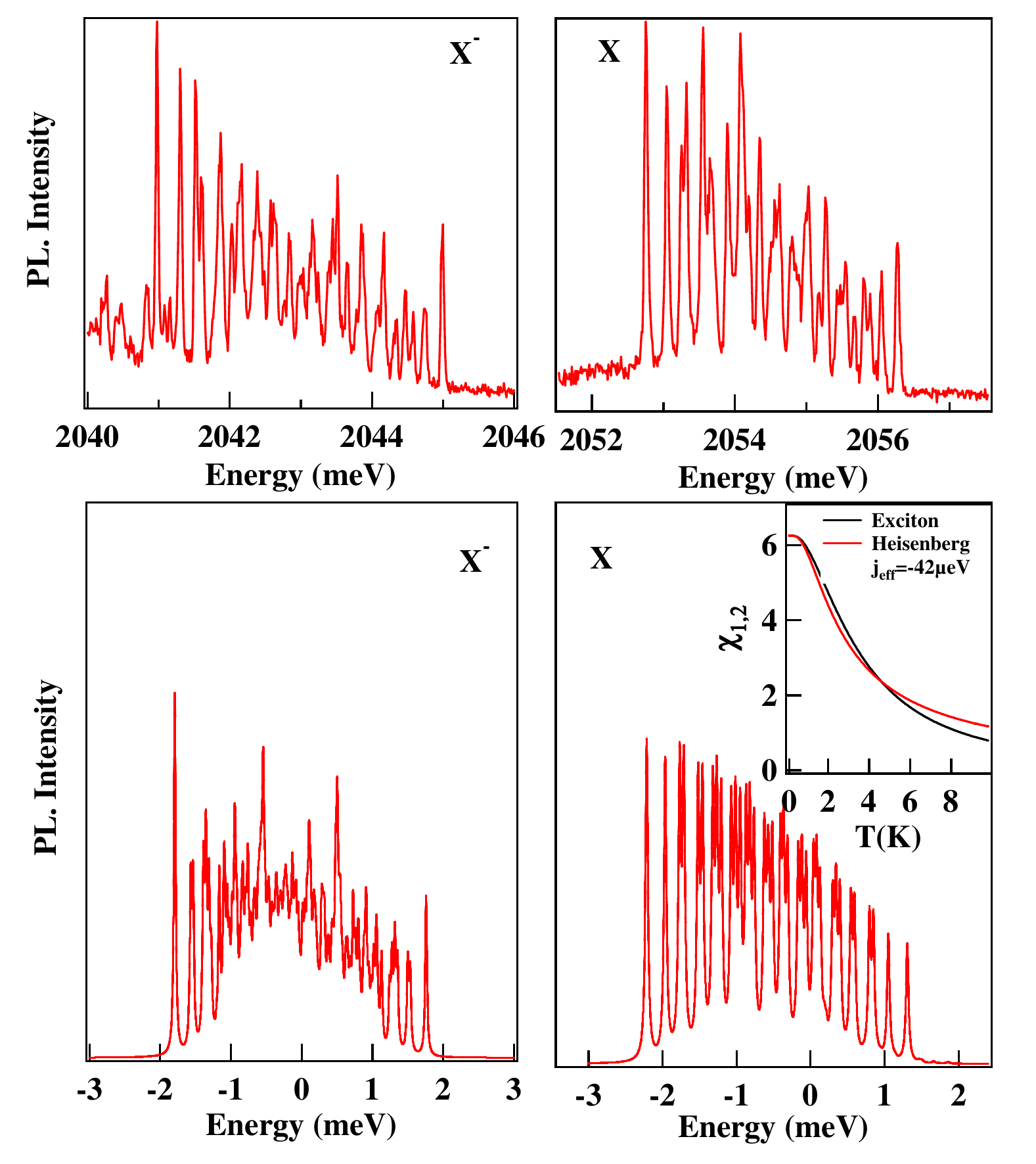}
\caption{Experimental (top panels) and calculated (bottom
panels) spectra of X and X$^-$ exchanged coupled to 2 Mn in
a QD. See text for details of the calculations performed
with I$_{e,1}$=-55, I$_{h,1}$=150, I$_{e,2}$=-90,
I$_{h,2}$=270, I$_{eh}=-600~\mu$eV, an effective valence
band mixing parameter $\rho/\Delta_{lh}$=0.025 and
$\theta=0$. A thermalization (effective temperature T=40K)
is included for the X-2Mn and hole-2Mn complexes. A
broadening of the lines of $40~\mu$eV is included in the
calculated spectra. The inset show the Mn-Mn correlation
function calculated for the model presented in the text
(exciton) compared with the spin-spin correlation function
for a Heisenberg model for two S=5/2.} \label{Figfit}
\end{figure}

The low temperature (T=5K) PL and the PL excitation (PLE)
spectra of a CdTe/ZnTe QD containing 2 Mn atoms are
presented in Fig.~\ref{fig1}. As in non-magnetic or singly
Mn-doped QDs, the emission of the $X$, $X^-$ and biexciton
($X^2$) is observed simultaneously \cite{Leger2007}. Each
excitonic complexe, X-2Mn, X$^-$-2Mn and X$^2$-2Mn, is
split by the exchange interaction with the 2 Mn spins. The
spin structure of this system is better illustrated by the
detailed spectrum of the X-2Mn presented in
Fig.~\ref{Figfit}. As each Mn sits in a different position
in the QD, the overlap between the carrier wave function
and each Mn atom is different. In this case, 36 PL lines
are expected for the X-2Mn complex. The higher and lower PL
lines correspond to the total spin projection of the 2 Mn
$M_z=\pm5$ coupled to the bright exciton ($\pm1$). The next
lines correspond to the situation where the less coupled Mn
spin has flipped its spin projection component by one unit
($M_z=\pm4$). The emission structure becomes more complex
as $M_z$ decreases (towards the center of the emission
structure) because different configurations of the 2 Mn
spins interacting with the exciton spin are very close in
energy.

The details of the PL are compared to a model based on an
effective  spin Hamiltonian that can be derived from the
$sp-d$ model for confined electrons and holes exchanged
coupled to the Mn spins \cite{Fernandez2006,Cao2011}.  The
spin interacting part of the Hamiltonian of a QD describing
the interaction of the electron spin $\vec{\sigma}$, the
heavy hole spin $\vec{J}$, and  two Mn  spins
$\vec{S}_{1,2}$ reads:

\begin{eqnarray}
{\cal H}&=&
\vec{\sigma}\cdot\left(I_{e,1} \vec{S_1}+
I_{e,2}\vec{S_2}\right)+\vec{J}\cdot\left(
I_{h,1}\vec{S_1}+
I_{h,2}\vec{S_2}\right)\nonumber\\
&+&I_{eh}\vec{\sigma}.\vec{J}+I_{12}\vec{S_1}.\vec{S_2}
\label{Hamilton1}
\end{eqnarray}

\noindent where the hole spin operator, represented in the
basis of the two low energy heavy hole states, are related
to the Pauli matrices $\tau$ by $J_{\pm}= \eta \tau_{\pm}$
and $J_z= \frac{3}{2}\tau_z$,with
$\eta=\frac{-2\sqrt{3}e^{-2i\theta}\rho}{2\Delta_{lh}}$,
$\rho$ the coupling energy between heavy-holes and
light-holes split by $\Delta_{lh}$ and $\theta$ the angle
relative to the [110] axis of the principal axis of the
anisotropy responsible for the valence band mixing
\cite{Leger2007}. I$_{h,i}$ (I$_{e,i}$) is the exchange
integral of the hole (electron) with the Mn atom $i$,
I$_{eh}$ the electron-hole exchange interaction and
$I_{12}$ the short ranged anti-ferromagnetic Mn-Mn
interaction. This latter coupling is only comparable with
the carrier-Mn energy when the 2 Mn are positioned closed
to each other ($I_{Mn,Mn}$=0.5meV for neighboring atoms
\cite{Furdyna1988,Qu2006}): it is neglected in the
following because the carrier-Mn interaction is different
for the two atoms indicating that they are far apart in the
QD.

The emission spectrum of the X$^-$-2Mn can be computed
using a similar spin effective Hamiltonian. It corresponds
to the optical transitions between an initial state where
the 2 Mn spins interact with a hole spin and two paired
electron spins and a final state where the remaining
electron is exchanged coupled with the 2 Mn spins. The
small energy correction produced by the Mn-Mn interaction
mediated by the presence of the two electrons in the
initial state \citep{Qu2006} is neglected here.

The calculated spectra are presented in Fig.~\ref{Figfit}.
The main feature ({\it i.e.} overall splitting, position of
PL lines, linear polarization) of the experimental spectra
can be well reproduced by this spin effective model. The QD
presented here corresponds to a situation where the
coupling with one Mn is about twice the coupling with the
other (the parameters are listed in the caption of
Fig.~\ref{Figfit}). To reproduce the intensity distribution
in the PL spectra, a thermalization on the X-2Mn levels
with an effective temperature T=40K is included in the
model. The same thermalization is used for the hole-2Mn
levels to reproduce the X$^-$-2Mn spectrum suggesting a
hole-2Mn spin relaxation during the lifetime of X$^-$.

For a given set of parameters, and a given charge state in
the dot, we solve numerically  the multi-spin Hamiltonian,
obtaining eigenstates $\Psi_n$ and their energies $E_n$ and
we can compute the spin correlation function of the 2 Mn:
\begin{equation}
\chi_{1,2}(k_BT)\equiv \langle
\vec{S}_1\cdot\vec{S}_2\rangle=\sum_n P_n(k_BT) \langle
\Psi_n| \vec{S}_1\cdot\vec{S}_2|\Psi_n\rangle
\end{equation}
where $P_n=e^{-E_n/k_BT}/(\sum_{n}e^{-E_n/k_BT})$ is for
the Boltzmann occupation function. The curve $\chi_{1,2}(k_
BT)$ is shown in the inset of Fig.~\ref{Figfit}, together
with the analogous curve for a Heisenberg model
$j_{eff}\vec{S}_1\cdot\vec{S}_1$. The $\chi_{12}$ for the
Heisemberg model depends only on the ratio  $j_{eff}/k_BT$
so that we   take $j_{eff}$ as the one that minimize
difference between the two correlation curves.  This
procedure permits to extract an energy to characterize the
indirect Mn-Mn correlation.  When we use this procedure for
the case of 1 electron Heisenberg coupled to 2 Mn, we find
$j_{eff}=-6\mu$eV,  for the parameters given in the caption
of figure 2 and a scaling of $j_{eff}
/\sqrt{I_{e,1}I_{e,2}}\simeq -0.05$, in agreement with the
effective coupling derived by one of
us\cite{Fernandez2004}.  For the same parameters, in the
case of the exciton mediated effective interaction,  we
find $j_{eff}=-42\mu$eV
 and a scaling
$ j_{eff}/\sqrt{I_{h,1}I_{h,2}} \simeq -0.2$.

Similarly to the singly Mn doped QDs\citep{Leger2006}, the
overall splitting of the X$^-$-2Mn observed experimentally
is larger than the one obtained with the values of exchange
integrals deduced from the calculation of the X-2Mn
spectra. This can be understood by the enhancement of the
hole-Mn overlap through the increase of the Coulomb
attraction of the hole by the two strongly confined
electrons.

\begin{figure}[hbt]
\includegraphics[width=3.3in]{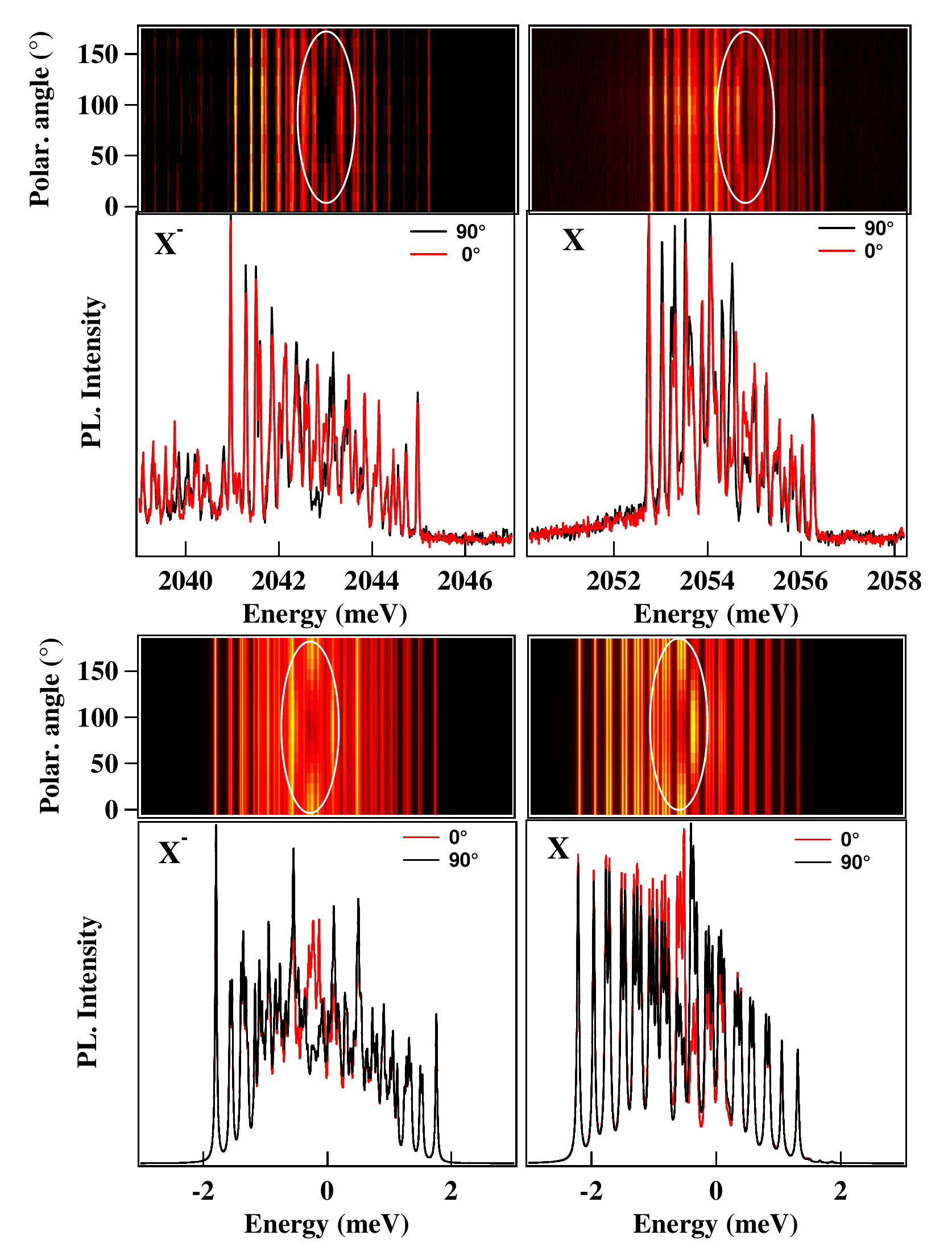}
\caption{Top panel: Experimental linear polarization
dependence of the PL intensity of X and X$^-$ exchanged
coupled to 2 Mn atoms. The polarization angle is relative
to the [110] direction. The PL are presented for orthogonal
linear analyzer directions (red and black curves). Bottom
panel: Calculated linear polarization dependence of the PL
intensity of X and X$^-$ with the same parameters as the
one given in Fig.\ref{Figfit}.}\label{fig3}
\end{figure}

Figure~\ref{fig3} reveals the linear polarization rate
which is observed in the center of the X-2Mn and X$^-$-2Mn
PL spectra. This polarization can be explained by the
presence of a valence band mixing. This mixing, combined
with the short range electron-hole exchange interaction,
couples the bright excitons +1 and -1 associated with the
same Mn spins configurations \cite{Leger2007}. As in
non-magnetic QDs, this mixing creates linearly polarized
eigenstates. This mechanism is more efficient in the center
of the X-2Mn PL structure where $M_z$ is smaller and the +1
and -1 exciton close in energy. In the absence of
electron-hole exchange interaction (situation of a
negatively charged QD) the valence band mixing allows
simultaneous hole-Mn flip-flops. These flip-flops couple
states of the hole-2Mn complex in the initial state of the
negatively charged exciton optical transition. These mixed
states are at the origin of the linear polarization in the
spectra of the charged exciton \cite{Leger2007}. The linear
polarization is well reproduced by the spin effective model
(Fig.~\ref{fig3} (bottom panel)) for both the X-2Mn and
X$^-$-2Mn if a weak effective valence band mixing
($\rho/\Delta_{lh}$=0.025) is introduced.

\begin{figure}[hbt]
\includegraphics[width=3.5in]{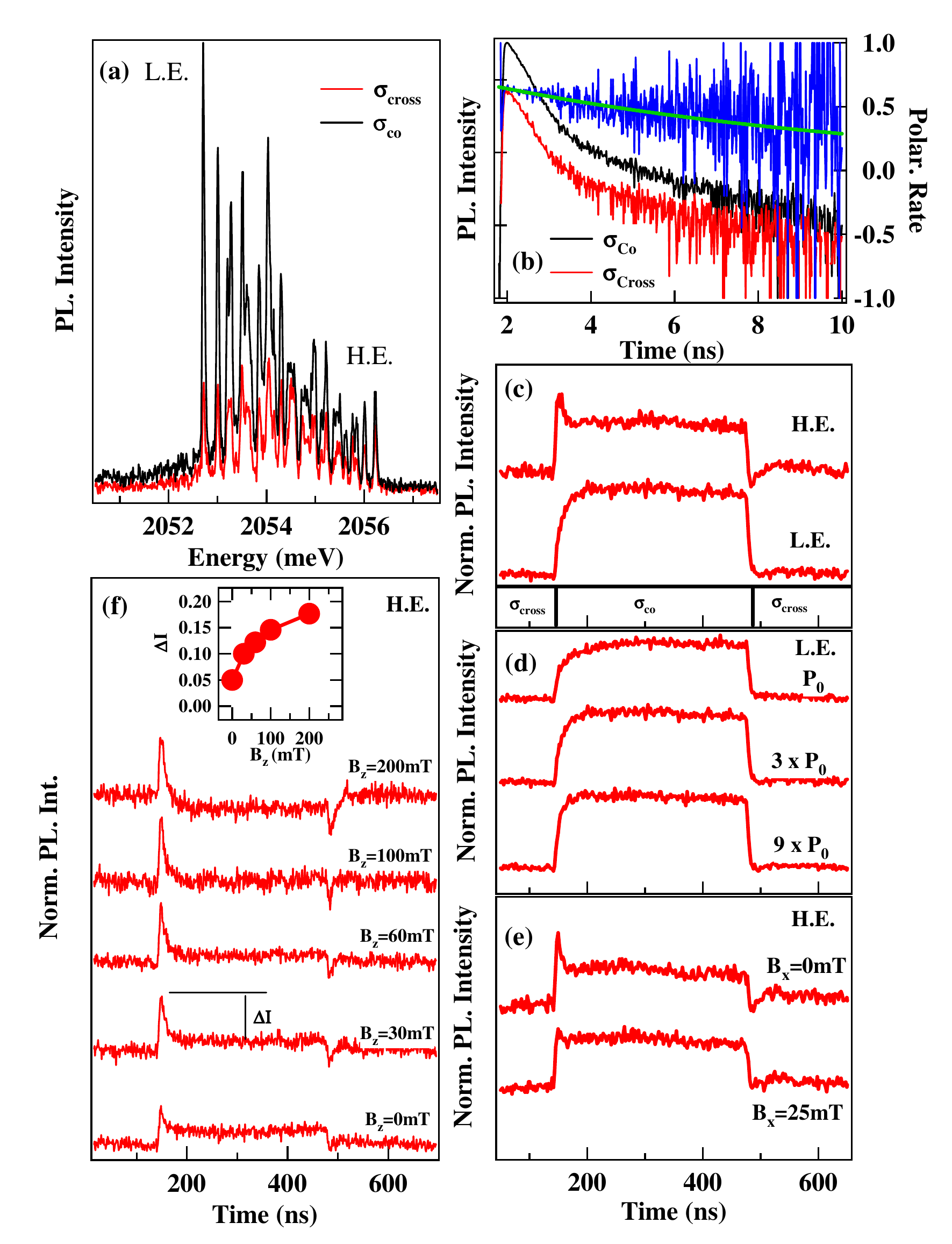}
\caption{(a) PL of X-2Mn under circularly polarized
excitation on an excited state (peak at 2143.5 meV in the
inset of Fig.~\ref{fig1}) and detected in co and cross
circular polarization. (b) Time resolved polarization rate
of the low energy line (L.E) under pulsed excitation at
2143.5 meV. (c) Optical orientation transients under
polarization switching of the excitation detected in
circular polarization on the low (L.E.) and high (H.E.)
energy lines. (d) Excitation intensity dependence of the
spin transients detected on the low energy line. (e) and
(f) present the magnetic field dependence of the spin
transients in Faraday (e) and Voight (f) configuration
detected on the high energy line.} \label{figpump}
\end{figure}

The 2 Mn spins can be optically oriented by the injection
of spin polarized carriers \cite{LeGall2009,LeGall2010}. To
optically pump the Mn spins, the QD is excited with a
tunable continuous wave laser on resonance with the excited
state corresponding to the strong PLE peak at 2143.5 meV
(Fig.~\ref{fig1}(b))\cite{Glazov2007}. The relative
intensity of the PL lines of the X-2Mn state depends
strongly on the correlation between the polarization of the
excitation and detection (Fig.~\ref{figpump}(a)). As each
line corresponds to a given $M_z$, this shows that the
whole process of spin injection and relaxation creates a
non-equilibrium distribution of the two Mn spin states. In
opposition to the observation in QDs containing a large
number of magnetic atoms \cite{klopotowski2011}, the
polarization of X-2Mn is almost fully conserved during the
lifetime of the exciton. Nevertheless, an exciton spin
relaxation time of about 10 ns can be extracted from the
time decay of the circular polarization rate of the exciton
(exponential fit in Fig.~\ref{figpump}(b)).

\begin{figure}[hbt]
\includegraphics[width=3.2in]{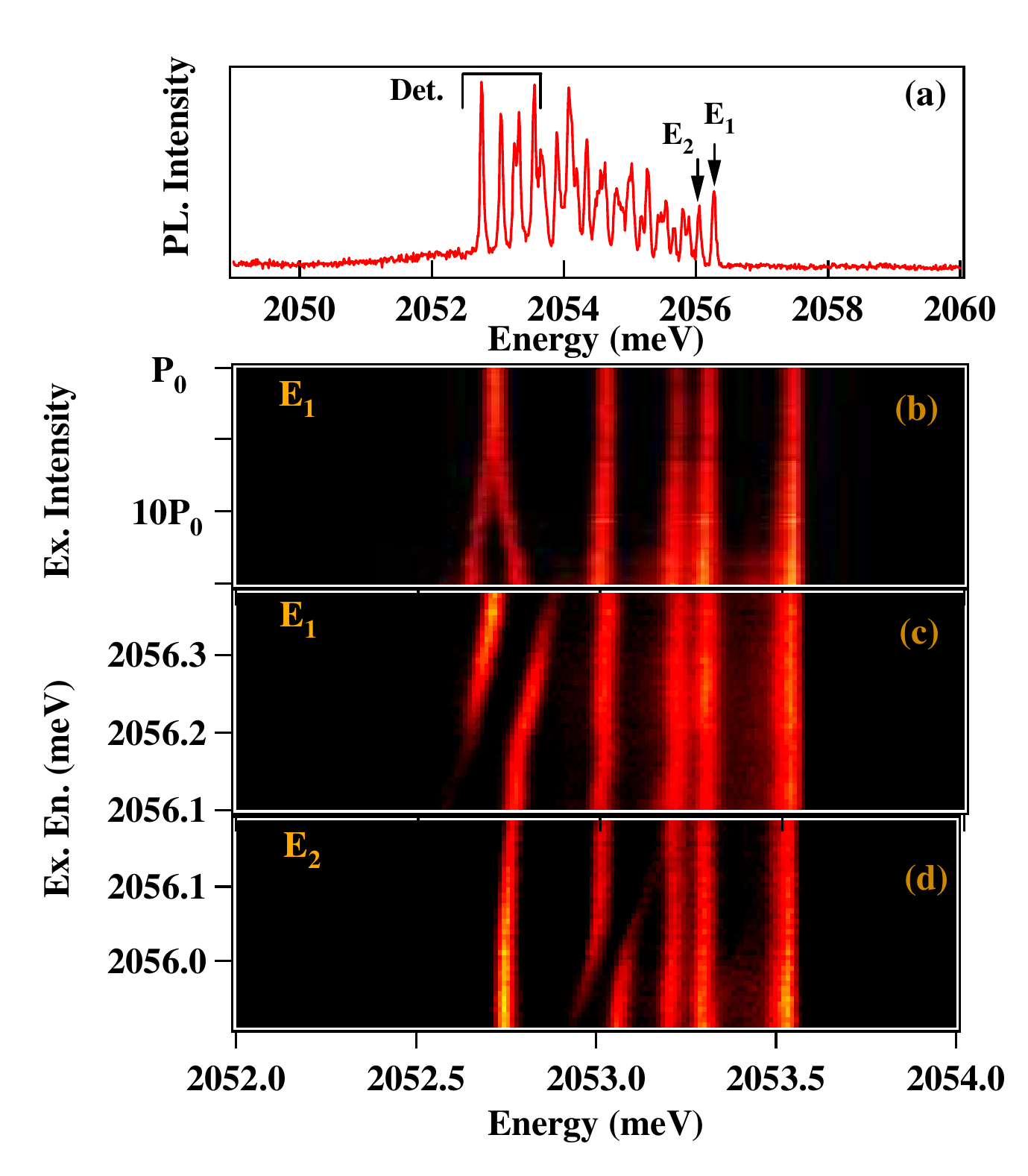}
\caption{(a) Nonresonant PL of X-2Mn.(b) PL map of the
X-2Mn versus the power of the control laser in resonance
with E$_1$. (c,d) PL maps of the X-2Mn when the control
laser energy is tuned around the 2-Mn spin states $M_z$=+5
(c) and $M_z$=+4 (d).} \label{fig5}
\end{figure}

The main features of time resolved optical orientation
experiments are reported in Fig.~\ref{figpump}(c-f) where
the polarization of the excitation laser is modulated
between two circular states by an electro-optic modulator
with a rise time of 10 ns combined with a quarter-wave
plate. Under these excitation conditions, switching the
circular polarization of the laser produces a change of the
PL intensity with two transients (Fig.~\ref{figpump}(c)):
first an abrupt one, reflecting the population change of
the spin-polarized excitons; then a slower transient with
opposite signs on the two extreme PL lines
(Fig.~\ref{figpump}(c)) ({\it i.e.}, when monitoring the Mn
spin states $M_z$=+5 or $M_z$=-5) and a characteristic time
which is inversely proportional to the pump intensity
(Fig.~\ref{figpump}(d)). This is the signature of an
optical orientation process which realizes a spin
orientation of the 2 Mn atoms.

As shown in Fig.~\ref{figpump}(f), the efficiency of the
spin orientation increases as soon as a magnetic field,
$B_z$, of a few tens of mT is applied in the Faraday
configuration \cite{LeGall2010}. By contrast, an in-plane
magnetic field ($B_x$) induces coherent precession of the
Mn spins away from the optical axis (QDs growth axis), so
that the average spin polarization, probed by the amplitude
of the optical pumping signal, decays \cite{LeGall2009}. A
transverse field of 25 mT is enough to completely erase the
optical pumping (Fig.~\ref{figpump}(e)).

Finally, it is shown in Fig.~\ref{fig5} that a selective
addressing of the X-2Mn complex can be achieved with a
control laser on resonance with one of the PL line and the
energy of the 2 Mn spins tuned optically \cite{LeGall2011}.
When a control single mode laser is tuned to the high
energy line of X-2Mn in $\sigma+$ polarization, a splitting
is detected in $\sigma-$ polarization on the low energy
line of X-2Mn. The control laser field mixes the states
with a 2Mn spin component $M_z$=+5 in the presence (X-2Mn)
or absence (2Mn alone) of the exciton. At resonance, hybrid
matter-field states are created (Fig.~\ref{fig5}(b)). As
the laser detuning increases, the optically active
transitions asymptotically approach the original excitonic
transitions where the remaining energy offset is the
optical Stark shift. The use of a resonant laser field
allows to individually address any spin state of the 2 Mn
if the corresponding excitonic transition is sufficiently
isolated from the others (Fig.~\ref{fig5}(c) and (d)).

In summary, we demonstrated the possibility to access the
spin state of 2 Mn atoms embedded in a CdTe/ZnTe QD. The
two spins can be optically oriented and their energy tuned
by a resonant laser field. As each of these spins is
exchanged coupled to the confined carriers, one would
expect to be able to control their mutual interaction. Each
Mn spin could be individually addressed by a resonant
microwave excitation while their coupling could be turned
on by the controlled injection of an individual carrier.

This work is supported by the French ANR contract QuAMOS,
Fondation Nanoscience (RTRA Grenoble) and EU ITN project
Spin-Optronics. JFR acknowledges funding from Ministerio de
Economia, grants FIS2010-21883-C02-01, Plan Nacional de
I+D, codigo FIS- No. FIS2010-21883-C02-01, and CONSOLIDER
CSD2007-0010, and from Generalitat Valenciana, grant
Prometeo 2012-011.

\end{document}